\documentclass[aps,pra,reprint,groupedaddress,amsmath,amssymb,showpacs,preprintnumbers]{revtex4-1}

\usepackage{graphicx}
\usepackage{dcolumn}
\usepackage{bm}
\usepackage{hyperref}
\usepackage{type1cm}
\usepackage[caption=false]{subfig}
\usepackage{enumitem}
\usepackage{color}

\newcommand{\rydstate}{$|r\rangle =\mathrm{5}s\mathrm{48}s\,^\mathrm{1}\!S_\mathrm{0}\ $}

\newcommand{\groundstate}{$|g\rangle=\mathrm{5}s^{\mathrm{2}}\,^{\mathrm{1}}\!S_{\mathrm{0}}\ $}

\begin{document}


\title{Radiation trapping in a dense cold Rydberg gas}


\author{D. P. Sadler}
\author{E. M. Bridge}
\author{D. Boddy}
\author{A. D. Bounds}
\author{N. C. Keegan}
\author{G. Lochead}
\author{M. P. A. Jones}
\email[]{m.p.a.jones@durham.ac.uk}
\affiliation{Joint Quantum Centre Durham-Newcastle, Department of Physics, Durham University, Durham, DH1 3LE, UK}
\author{B. Olmos}
\affiliation{Centre for the Mathematics and Theoretical Physics of Quantum Non-equilibrium Systems, The University of Nottingham, Nottingham, NG7 2RD, UK}
\affiliation{School of Physics and Astronomy, University of Nottingham, Nottingham NG7 2RD, UK}


\date{\today}


\begin{abstract}
Cold atomic gases resonantly excited to Rydberg states can exhibit strong optical nonlinearity at the single photon level. We observe that in such samples radiation trapping leads to an additional mechanism for Rydberg excitation. Conversely we demonstrate that Rydberg excitation provides a novel {\em in situ} probe of the spectral, statistical, temporal and spatial properties of the trapped re-scattered light. We also show that absorption can lead to an excitation saturation that mimics the Rydberg blockade effect. Collective effects due to multiple scattering may co-exist with co-operative effects due to long-range interactions between the Rydberg atoms, adding a new dimension to quantum optics experiments with cold Rydberg gases.
\end{abstract}

\pacs{}

\maketitle


\section{Introduction}
The study of collective effects in light scattering is an important application of ultra-cold atomic gases. Multiple scattering plays a key role in laser cooling, where it limits the phase space density of the magneto-optical trap (MOT) \cite{Walker1990}. The reduction in Doppler broadening also facilitates studies of effects such as coherent backscattering  \cite{Labeyrie1999}, weak localization \cite{Jonckheere2000,Kupriyanov2005} and random lasing \cite{Baudouin2013}. More recent experiments have probed coherent collective effects in free induction decay \cite{Kong2014}. In such experiments divalent atoms provide a significant advantage, due to the absence of hyperfine structure \cite{Bidel2002,Kong2014} and the presence of narrow intercombination lines.

As the density is increased, the separation between the atoms can become smaller than the optical wavelength and co-operative effects due to dipole-dipole interactions become important \cite{Dicke1954}. Recent results in this context include observations of the co-operative Lamb shift \cite{Rohlsberger2010,Keaveney2012} and suppressed transverse scattering \cite{Pellegrino2014,Bromley2016} as well as proposals using divalent atoms to study interacting many-body systems \cite{Olmos2013,Zhu2015,Bettles2015}.

Co-operative nonlinear effects can be induced at lower densities by coupling an optical transition to a high-lying Rydberg level \cite{Pritchard2010}. Here, the strong long-range interactions between Rydberg atoms lead to a blockade effect \cite{Lukin2001} that restricts the number of excited atoms in a given spatial region. Mapping this blockade onto an optical transition under electromagnetically induced transparency (EIT) conditions \cite{Mohapatra2007,Mauger2007} results in a medium with non-linear absorption at the single-photon level \cite{Dudin2012,Peyronel2012,Maxwell2013}, enabling recent demonstrations of single-photon transistors \cite{Gorniaczyk2014,Tiarks2014}, impurity imaging \cite{Gunter2012,Olmos2011} and proposals for photonic quantum logic gates \cite{Friedler2005,Gorshkov2011,Paredes2014,Khazali2015}. In all of these works the medium was required to have a high optical depth since all unwanted photons must be absorbed \cite{Peyronel2012}. Therefore multiple scattering might be expected to also play a role, though its effect has not yet been discussed.
\begin{figure}
\includegraphics[width=1\columnwidth]{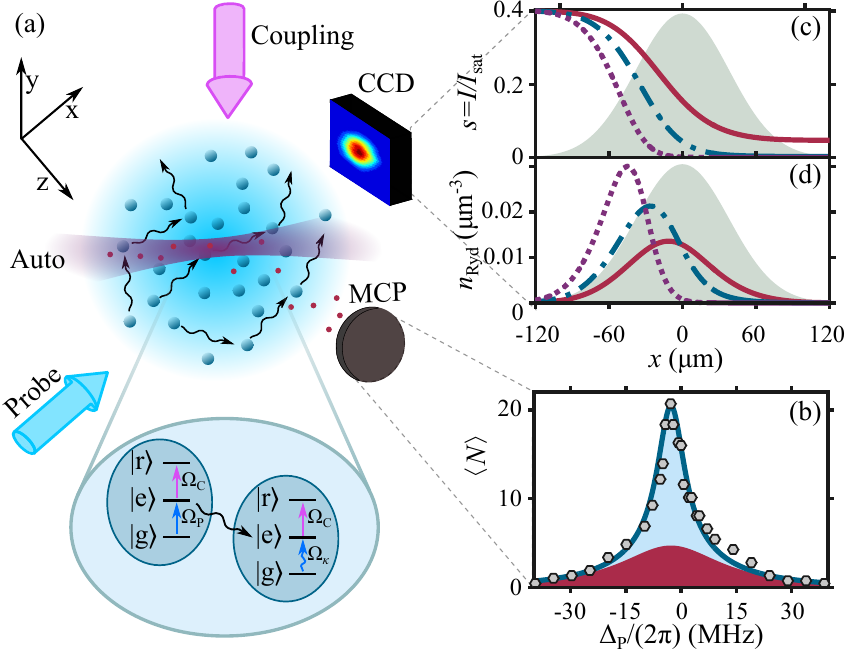}
\caption{(Color online) (a) A dense cloud of ground state \groundstate Sr atoms is excited by probe and coupling lasers to the Rydberg state \rydstate via the intermediate state $|e\rangle=\mathrm{5}s\mathrm{5}p\,^{\mathrm{1}}\!P_{\mathrm{1}}$. An autoionization beam provides spatially-resolved Rydberg detection. Rydberg atoms are also excited by rescattered probe light trapped in the cloud leading to (b) a broad component in the excitation spectrum (red) in addition to the narrow laser-excited component (blue). (c) Increasing optical depth ($b=3$ (red), $6$ (blue) and $14$ (purple)) leads to attenuation of the probe beam (relative intensity $s$) with propagation distance $x$, leading to (d) the confinement of laser-excited Rydberg atoms (density $n_{\mathrm{Ryd}}$) to the edge of the cloud (shaded area). \label{fig:fig1}}
\end{figure}

In this paper we demonstrate that multiple scattering can play an important role in dense, optically thick Rydberg gases. Radiation trapping leads to a significant additional population of Rydberg atoms within the cloud (Fig. \ref{fig:fig1}), modifying both the spectrum and statistics of the Rydberg excitation process. Conversely, we show that Rydberg excitation provides a probe of the spectral, spatial and temporal properties of the trapped light which operates {\emph{in situ}}, rather than relying on measurements of the light that escapes. We observe a saturation of the number of Rydberg atoms with increasing density that mimics the effect of the Rydberg blockade, but which in fact occurs due to optical depth effects \cite{Hofmann2013,Garttner2013}. Finally, we discuss the possible co-existence of collective effects due to both multiple scattering and long-range Rydberg-Rydberg interactions. 


\section{Experimental setup}\label{sec:exp}
A schematic of the experiment is shown in Fig.~\ref{fig:fig1}(a). A cloud of up to $10^6$ $^{88}$Sr atoms at $<10\,\mu$K was prepared using a two-stage MOT \cite{Boddy2014}. The atoms were subsequently excited for a duration $\tau$ by probe ($|g\rangle\to|e\rangle$) and control ($|e\rangle\to|r\rangle$) laser beams (both larger than the cloud) with wavelength and Rabi frequencies $\lambda_\mathrm{P}=461$\ nm, $\Omega_\mathrm{P}/(2\pi) = 10$\ MHz and $\lambda_\mathrm{C}=413$\ nm, $\Omega_\mathrm{C}/(2\pi) = 0.6$\ MHz, respectively. 
The probe beam was linearly polarized orthogonal to the propagation direction of the circularly polarized coupling beam. Since the intermediate state $|e\rangle$ decays quickly ($\Gamma_\mathrm{P} = 2\times 10^8\ \mathrm{s}^{-1}$), $\Gamma_\mathrm{P}>\Omega_\mathrm{P}>\Omega_\mathrm{C}$ such that the experiment operated in the strong probe \cite{Wielandy1998} and strongly dissipative regime \cite{Ates2006,Ates2007,Lesanovsky2013,Schempp2014,Malossi2014,Urvoy2015}, where the effect of EIT on the probe beam propagation \cite{Datsyuk2006} is less than 1\%.

Detection of the Rydberg atoms was carried out by applying a 2\,$\mu$s autoionisation pulse of light resonant with the Sr$^{+}$ D2 transition at 408\,nm \cite{Lochead2013}. The autoionization beam propagated at $30^\circ$ to the probe beam and was focused to a $1/e^2$ waist of approximately 6 $\mu$m. It was aligned with the cloud center by measuring the ion signal as a function of position as shown in Fig.~3(b).
A voltage pulse applied to split-ring electrodes surrounding the cloud directed the resulting ions towards the micro-channel plate (MCP). Voltage pulses corresponding to individual ions were counted using a fast digital oscilloscope. After 100 repetitions for each set of experimental conditions we obtained the mean number of detected ions $\langle N \rangle$ and Mandel $Q$-parameter $Q=((\langle N^2\rangle - \langle N \rangle^2)/\langle N \rangle)-1$. We have checked carefully for detector saturation and found that it affects the results only when $\langle N \rangle \gtrsim 60 $. A small background due to spontaneous ionization is detected and removed \cite{Millen2010}.

Absorption imaging on the probe transition was used to obtain the cloud size along the propagation and gravity directions ($x_0$ and $y_0$ respectively), as well as the peak optical depth  $b=\max_{(y,z)}\{-\ln[I_\mathrm{T}(y,z)/I_\mathrm{0}(y,z)]\}$, where $I_\mathrm{T}(y,z)$ and $I_\mathrm{0}(y,z)$ are the transverse distributions of the transmitted and incident intensity respectively. Since the cloud was optically thick, these parameters were extracted using a Gaussian fit to the wings of the cloud. The cloud was assumed to be symmetric around the $y$-axis, such that the cloud size along the $x$-axis ($x_0$) is equal to that along the $z$-axis ($z_0$). We quote the statistical uncertainty in the mean cloud size (standard error). To check for a systematic error due to the high optical thickness, we verified that the atom number obtained from the fit agreed with that measured after ballistic expansion at $b<1$. Fluctuations in the atom number dominate the uncertainty in the density (optical thickness), which is $\sim 10\%$.

\section{Excitation spectrum}\label{sec:spectra}
\begin{figure}
\includegraphics[width=1\columnwidth]{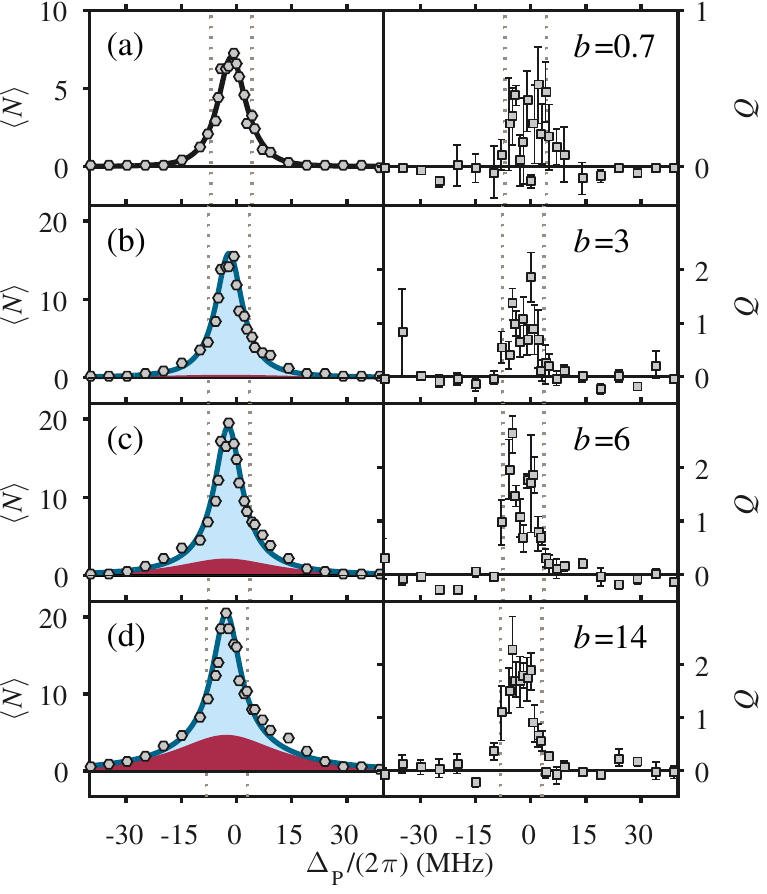}
\caption{(Color online) (Left) Mean number of detected ions $\langle N \rangle$ and (Right) $Q$-parameter as a function of  $\Delta_{\mathrm{P}}$. Gray dashed lines indicate $\pm W_\mathrm{L}$. (a) $b=0.7$, solid black line is the OBE model. (b) $b=3$ (c) $b=6$, and (d) $b=14$ (peak ground state density $n_0=1.5\times 10^{12}$~cm$^{-3}$). Solid blue lines show the two-component OBE model, and red dashed lines indicate the pedestal component. Here $x_0=89\pm3\ \mu$m. \label{fig:fig2}}
\end{figure}

The excitation spectrum was obtained by varying the probe laser detuning $\Delta_{\mathrm{P}}$ while the coupling laser frequency was fixed on resonance. At low $b$  (Fig.~\ref{fig:fig2}(a)), the spectrum is well described by the solution of the optical Bloch equations (OBE) for non-interacting atoms. To correctly account for the laser polarization we explicitly include the three $m_J$ sublevels of the intermediate state, leading to a 5-level model (see Appendix A). The observed linewidth (FWHM) $W_\mathrm{L} = 4\ $MHz is largely determined by technical noise on the excitation lasers. In previous work  \cite{Lochead2012,Lochead2013} we have shown that the inclusion of the measured technical noise in the OBE model also leads to a quantitative explanation of the observed super-Poissonian $Q>0$ excitation statistics.

As the cloud becomes optically thick (Fig.~\ref{fig:fig2}(b--d)) a broad pedestal appears, which grows in amplitude as $b$ increases. The pedestal exhibits approximately Poissonian statistics  ($Q\approx 0$), with the region of super-Poissonian statistics remaining confined to the narrow central feature. Slow fluctuations in the laser detunings can cause a shift of ($\sim2$ MHz) of the narrow feature relative to the (fixed) zero detuning point. Within this uncertainty, both features remain centered on resonance.



The results in Fig.~\ref{fig:fig2} indicate the emergence of a new mechanism for Rydberg excitation at high optical depth, with different spectral and noise characteristics from the direct laser excitation. We attribute the appearance of the pedestal to multiple scattering. At $b=14$, the optical mean free path is $l=x_0/b = 6\,\mu$m. The probe laser is thus rapidly attenuated (see Fig.~\ref{fig:fig1}(c)) and Rydberg excitations are only created by the probe laser on the edge of the cloud (see Fig.~\ref{fig:fig1}(d)). In contrast, scattered probe photons undergo up to  $b^2 =200$ scattering events before leaving the cloud and become trapped. The pedestal appears because multiple scattering modifies the spectral distribution of the trapped probe light, broadening and shifting it towards the line center. The trapped light combines with the coupling laser (for which the cloud remains transparent) to excite additional Rydberg atoms. Frequency redistribution also leads to the observed reduction in the $Q$-parameter, since the spectrum of the multiply-scattered light decouples from that of the probe laser.



\section{Two-component model}\label{sec:model}
We have constructed a model (see Appendix A) for the data in Fig.~\ref{fig:fig2}(d) based on the hypothesis that the spectrum consists of two distinct components originating from laser excitation and multiple scattering. We justify this approach in two ways. Firstly,  the $Q$-parameter strongly suggests different excitation mechanisms for each part of the spectrum. Secondly, the rapid attenuation of the probe beam means that laser excited Rydbergs are created at the edge of the cloud, whereas multiple scattering should favor the central region of the cloud where the density is highest. Therefore the overlap region where Rydberg excitation occurs due to both mechanisms is small. We also treat the atoms as non-interacting, which provides a direct test of the hypothesis that the behavior that we observe is primarily due to multiple scattering.

Under these hypotheses, the shape of the laser excited (narrow) component is independent of $b$ \cite{propnote} and can be modeled using an appropriately frequency shifted and amplitude scaled version of the solid curve in Fig.~\ref{fig:fig2}(a). To model the pedestal we also use the optical Bloch equations, with the probe laser replaced by the trapped radiation field. To do so we must make a number of assumptions about the spectrum of the trapped light. Firstly, we assume that the spectrum of the light trapped within the cloud evolves to a steady state on a timescale on the order of the photon trapping time
\[
t_\mathrm{trap}\approx \frac{3}{\alpha \pi^2}\frac{b^2}{\Gamma_\mathrm{P}}\approx 60\ \mathrm{ns},
\]
where $\alpha=5.35$ for a spherical Gaussian cloud \cite{Labeyrie2003}, such that it can be treated as constant during the excitation pulse. Secondly, we assume that the steady-state bandwidth of the trapped light is set by the Lorentzian absorption coefficient of atoms at rest, since light outside this bandwidth is more likely to escape. Under these assumptions, all of the multiply-scattered photons have the same spatio-temporal envelope, determined by $\Gamma_\mathrm{P}$ plus any additional power broadening. Therefore, we include the spectral width of the rescattered field as an extra homogeneous broadening term within the optical Bloch equations \cite{Dalton1992}.

Though Doppler broadening has been shown to play a role in multiple scattering even for cold atoms \cite{Labeyrie2003,Labeyrie2005,Pierrat2009}, we find that it remains negligible for our parameters even taking into account measured recoil heating during the probe pulse. Instead, rapid frequency redistribution occurs via spontaneous emission  \cite{Walker1990}, ``filling in'' the available trapping bandwidth. Since the trapped light may be incident from any direction we assume that it is unpolarized, coupling equally to all the Zeeman sublevels of the intermediate state. The remaining unknown is the detuning-dependent intensity of the trapped light, which strongly influences the shape of the pedestal. We assume that the intensity is proportional to the power absorbed from the laser, which is determined by the Lorentzian absorption coefficient \cite{intennote}. The Rabi frequency associated with the trapped probe field at line center ($\Delta_\mathrm{P}=0$) $\Omega_\kappa$ and an amplitude scaling factor are treated as fit parameters.

The resulting two-component OBE model is shown in Fig.~\ref{fig:fig2}(b-d). The model reproduces the main features of the spectra across roughly an order of magnitude in optical depth. However close inspection reveals that the model overestimates the wings of the data and empirically we find slightly better agreement using a Gaussian lineshape for the pedestal. The fit to the data in Fig.~\ref{fig:fig2}(d) yields $\Omega_\mathrm{\kappa}/(2\pi) = 8$ MHz, which is comparable to that of the probe laser, emphasizing the importance of the trapped radiation field and the importance of inelastic scattering effects.
Concerning the $Q$-parameter, the non-interacting one-body density matrix approach predicts Poissonian excitation statistics, in agreement with the data for the pedestal in Fig.~\ref{fig:fig2}. However, super-Poissonian statistics can result if classical fluctuations in the parameters (Rabi frequency, detuning) are present, as observed in the narrow component of the spectrum. The absence of excess noise for the pedestal therefore indicates that the spectrum of the multiply scattered light has decoupled from the classical fluctuations in the excitation lasers. This is expected, since significant frequency redistribution occurs even for a single inelastic scattering event. Sub-Poissonian statistics could also be observed, but cannot be derived from a one-body approach since correlations between the excitation probability of individual atoms are required. The minimum $Q$ that can be observed is bounded by the detection efficiency (5-10\%), so within statistical uncertainty the $Q$ values in  Fig.~\ref{fig:fig2} provide no evidence for correlated excitation due to Rydberg blockade.

\section{Excitation saturation and spatial dependence}\label{sec:sat}
\begin{figure}
\includegraphics[width=1\columnwidth]{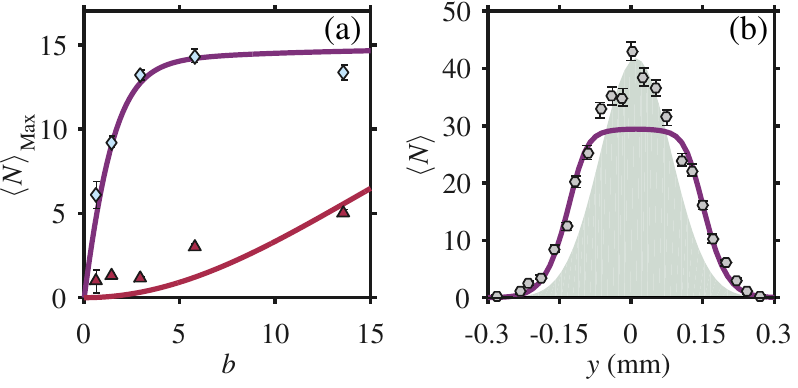}
\caption{(Color online) (a) $\langle N \rangle_\mathrm{Max}$ versus $b$ for the narrow (blue diamonds) and  pedestal (red triangles) components for the data in Fig.~\ref{fig:fig2} ($x_0=89\ \mu$m). Solid lines are the propagation model for the narrow component (purple) and the OBE model for the pedestal (red).  (b) $\langle N \rangle$ at $\Delta_\mathrm{P}=0$ versus $y$ (gray circles) and corresponding ground state density profile (shaded area) for a cloud with $x_0=380\pm4\ \mu$m and $b=8.4$ ($n_0=2\times 10^{11}$~cm$^{-3}$). Solid purple line is the propagation model fitted to the data at $|y|>$0.1\ mm.\label{fig:fig3}}
\end{figure}

Using the two-component model \cite{pedref} we extract the amplitude $\langle N \rangle_\mathrm{Max}$  (Fig.~\ref{fig:fig3}(a)) of both spectral components for the data shown in Fig.~\ref{fig:fig2}. The amplitude of the pedestal increases steadily with increasing $b$. To check if this behavior is captured by the OBE model, we assume that $\Omega_\kappa \propto b$ and fix the amplitude scaling factor to that obtained from Fig.~\ref{fig:fig2}(d). The resulting prediction for the variation of pedestal amplitude with $b$ is in reasonable agreement with the data.

The amplitude of the narrow component, shown in Fig.~\ref{fig:fig3}(a), saturates rapidly as $b$ increases. Saturation of the number of Rydberg excitations is often interpreted as a signature of the Rydberg blockade \cite{Tong2004,Singer2004}. However, in an optically thick cloud saturation may also occur due to attenuation of the probe laser, since the increase in density is canceled by a concomitant reduction in the illuminated volume of the cloud as shown in Fig.~\ref{fig:fig1}(c).

We quantitatively model this effect by solving the OBE model (section \ref{sec:model}) for laser excitation as a function of position in the cloud, obtaining the spatial distribution of the probe beam intensity and the density of Rydberg excitations. The 5-level OBE model for the narrow component is solved to obtain the dependence of the absorption cross-section $\sigma(I)$ and Rydberg population $\rho_{rr}(I)$ on the probe beam intensity $I$. Arrays are used to represent the  ground state density distribution $n(x,y,z)$, intensity $I(x,y,z)$ and Rydberg state population $\rho_{rr}(x,y,z)$. The intensity in the $j$th $y-z$ plane is calculated iteratively using the Beer-Lambert law, taking into account the intensity-dependent cross section. The corresponding density distribution of Rydberg atoms is given by
$n_{\mathrm{Ryd}}(x,y,z)=n(x,y,z) \rho_{rr}(I(x,y,z))$.
Finally, the number of ions is obtained by integrating the result of the model over a volume that approximately corresponds to that of the autoionization beam, such that
\[
N_{\mathrm{Ryd}}(Y,Z)=\sum_{x=-L}^{L}\sum_{y=(Y-W)}^{Y+W}\sum_{z=(Z-W)}^{Z+W}n_{\mathrm{Ryd}}(x,y,z)\,\mathrm{d}x^3,
\]
where $W$ is the size of the autoionization beam in units of the grid cell size $dx$ and $2L$ is the length of the array in the $x$ direction. The angle between the probe and autoionization beams is neglected.

Example results are shown in Fig.~\ref{fig:fig1}(d), which clearly show how increasing optical depth confines laser-excited Rydberg atoms to the low-density wings of the cloud. By integrating the model over a detection volume that represents the autoionization beam, we obtain a prediction for the variation of the amplitude of the narrow component with $b$, which is compared to the data in Fig.~\ref{fig:fig3}(a). The only fit parameter is an overall detection efficiency $\eta=0.06$. The curve is in very good agreement with the data, indicating optical depth and not Rydberg blockade is the dominant effect responsible for the saturation we observe. Therefore in order to unambiguously observe Rydberg blockade in the optically thick regime, the full statistical distribution of excitations is required. Sub-Poissonian statistics then provides access to the blockade-induced correlations. A similar conclusion was reached in experiments with Rydberg dark-state polaritons \cite{Hofmann2013,Garttner2013}.

The quantitative agreement of the propagation model with the data in Fig.~\ref{fig:fig3}(a) enables us to study the spatial distribution of Rydberg atoms created by radiation trapping. By translating the autoionization beam relative to the cloud along the $y$-axis, we measured the dependence of the total Rydberg signal on position \cite{Lochead2013} (Fig.~\ref{fig:fig3}(b)). The spatial distribution is clearly not in agreement with the shape of the ground state density distribution (shaded area). The propagation model provides a quantitative prediction of the spatial distribution of laser-excited atoms, which agrees well with the wings of the cloud but that is flat in the center. The flattening occurs because optical depth causes the number of Rydberg atoms to saturate in the dense central region of the cloud, but not in the wings. In contrast, the data shows no flattening, instead displaying a clear excess of Rydberg excitations in the center of the cloud. We attribute these extra Rydberg excitations to radiation trapping, and as might be expected they are concentrated in the dense central region where the trapped intensity is the highest.

The time evolution of the excitation spectrum is shown in Figs.~\ref{fig:fig4}(a-b). A pedestal is apparent even for the shortest time we can probe experimentally and the shape of the pedestal does not change as $\tau$ increases. This supports our assumption that fast redistribution occurs on the scale of $t_\mathrm{trap}$. The temporal evolution of the two components is shown in Fig.~\ref{fig:fig4}(c), along with data taken at low $b$  where the pedestal was negligible. The pedestal grows with $\tau$, since its amplitude is governed by the slow timescale associated with the coupling laser, and its behavior is in reasonable agreement with the OBE model for $\tau< 4\ \mu$s. The narrow component agrees very well with the OBE model at low-$b$ but the model cannot be scaled to reproduce the evolution of the narrow component at high-$b$, with the data rising faster and saturating more quickly than the model predicts.

\section{The effect of interactions}\label{sec:mc}
\begin{figure}
\includegraphics[width=1\columnwidth]{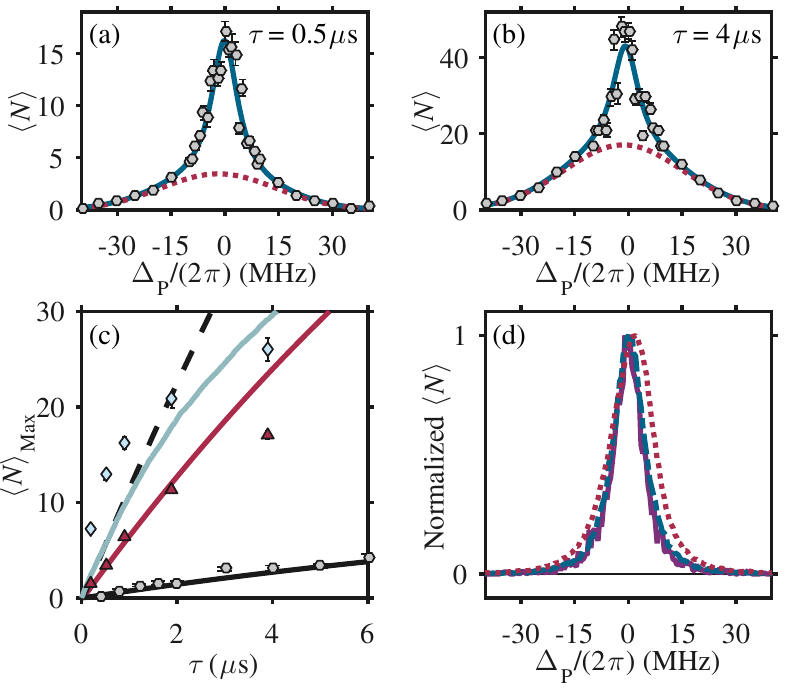}
\caption{(Color online)  $\langle N \rangle$ as a function of $\Delta_\mathrm{P}$ for $\tau= 0.5\ \mu$s (a) and $4\ \mu$s (b). Here $x_0=350\pm4\ \mu$m and $b=17$ ($n_0=4\times 10^{11}$~cm$^{-3}$). Solid lines are two-component fits with a Gaussian pedestal. (c) $\langle N\rangle_\mathrm{Max}$ versus $\tau$ for the narrow (blue diamonds) and pedestal (red triangles) components. Gray circles are $\langle N\rangle_\mathrm{Max}$ (narrow) at low $b$. Black solid and dashed lines indicate fits of the OBE model to the narrow component data at low and high $b$ respectively. The red solid line is the pedestal OBE model scaled to fit the data. The blue solid line is the result of the MC simulation. (d) Normalised $\langle N \rangle$ versus $\Delta_\mathrm{P}$ from the MC simulations, for densities $5 \times 10^9$ cm$^{-3}$ (yellow), $5 \times 10^{10}$ (blue), $1 \times 10^{11}$ cm$^{-3}$ (purple), $5 \times 10^{11}$ cm$^{-3}$ (red). \label{fig:fig4}}
\end{figure}

To investigate whether these effects could be due to interactions between Rydberg atoms, which we have so far neglected, we modeled the many-body excitation dynamics via a classical Monte-Carlo (MC) method. This method has been shown to correctly reproduce the dynamics of the quantum system in the regime $\Gamma_\mathrm{P}>\Omega_\mathrm{P}>\Omega_\mathrm{C}$ as under these conditions the dynamics of the coherences can be adiabatically eliminated \cite{Ates2007,Olmos2014,Marcuzzi2014}. Here we work in a three-level approximation where the degeneracy of the intermediate state is ignored, and the time-dependent model is solved numerically for a uniform (random) spatial distribution of atoms at each density. Multiple scattering is not included, so the results may only be compared to the narrow component of the spectrum. The interaction is included as a van der Waals-type $V(R)= C_6/R^6$, using the $C_6$ coefficient from \cite{Vaillant2012}. We note that the predicted blockade radius associated with the narrow component is $R_{\mathrm{B}}=(2C_6/\hbar W_{\mathrm{L}})^{1/6}\approx 3.4\ \mu$m. To find the appropriate density for comparison with the experiment we first solve the propagation model to find the ground state density that corresponds to the peak number of Rydberg excitations (see Fig.~\ref{fig:fig1}(d)). Qualitatively, the MC model is in better agreement with our data (Fig.~\ref{fig:fig4}(c)), although it still fails to reproduce the strong saturation at longer pulse lengths.The measured spontaneous ionization signal (10 counts 4~$\mu$s) indicates that we remain below the threshold for the creation of an ultra-cold plasma \cite{Killian1999,Millen2010} even for the data in Fig.~\ref{fig:fig4}(c).

The MC prediction for the spectrum is shown in Fig.~\ref{fig:fig4}(d). The result is in very good agreement with the narrow component observed in the data and the non-interacting OBE model for densities up to $10^{11}$ cm$^{-3}$.  At higher density, the line shifts and broadens asymmetrically, as expected from the van der Waals interaction potential. The reason this is not observed in our experiment is provided by the propagation model. As shown in Fig.~\ref{fig:fig1}(d), even for the highest density in Fig.~\ref{fig:fig2}(d), Rydberg excitation mostly occurs at densities below  $5\times10^{11}$~cm$^{-3}$ due to the attenuation of the probe beam. Therefore we expect the shape of the narrow component to show little density dependence, in agreement with our observations in Fig.~\ref{fig:fig2}.

In summary, our simple non-interacting two-component model reproduces most of the features of our data except at the highest optical depths. In principle, a more  complex  model that includes multiple scattering together with propagation and interactions \cite{Garttner2013} could be constructed using Monte-Carlo methods \cite{Avery1967,Garttner2013}. However modelling multiple scattering in the inelastic regime is very challenging, since the atomic absorption coefficient, resonance fluorescence spectrum and coupling to the Rydberg state are all intensity dependent \cite{Ellinger1994,Gremaud2006,Pohl2006,Wellens2010}, leading to a complex, spatially-dependent system that must be solved self-consistently. Such a model would be a powerful tool for studying multiple scattering in a gas of interacting scatterers, but is beyond the scope of this paper. 

\section{Conclusions}
In conclusion, we have shown that  multiple scattering plays an important role in dense Rydberg gases. It is likely to affect quantum devices based on Rydberg-mediated optical nonlinearities. Although photonic gates and single-photon sources operate with weak quantum probe fields, the possibility of creating unwanted additional Rydberg excitations could lead to dephasing or decoherence. For optical transistors that use a single photon to switch a strong classical pulse, multiple scattering may be a significant parasitic effect. Conversely we show that Rydberg excitation makes an excellent local probe of multiple scattering, providing information on the spectral, statistical, spatial and temporal characteristics of the trapped light within the cloud. This could enable the observation of exotic effects such as photon bubble formation \cite{Mendon2012,Rodrigues2016}. Our results suggest that while optical depth effects dominated the behavior, Rydberg-Rydberg interactions may still play a role. By varying the principal quantum number the relative importance of multiple scattering and co-operative nonlinearities due to interactions can be tuned, opening a new regime for optics experiments in cold gases.

\section*{acknowledgments}
The authors would like to thank I.~Hughes and C.~Adams for useful discussions. Financial support was provided by EPSRC grant EP/J007021/ and EU grants FP7-ICT-2013-612862-HAIRS and H2020-FETPROACT-2014-640378-RYSQ. B. Olmos was supported by the Royal Society and EPSRC grant DH130145. The data presented in this paper are available at https://doi.org/10.15128/dz010q04x.  

\appendix

\section{5-level optical Bloch equation model}
\begin{figure}
\centering
\vspace{0.5cm}
\includegraphics[width=1\columnwidth]{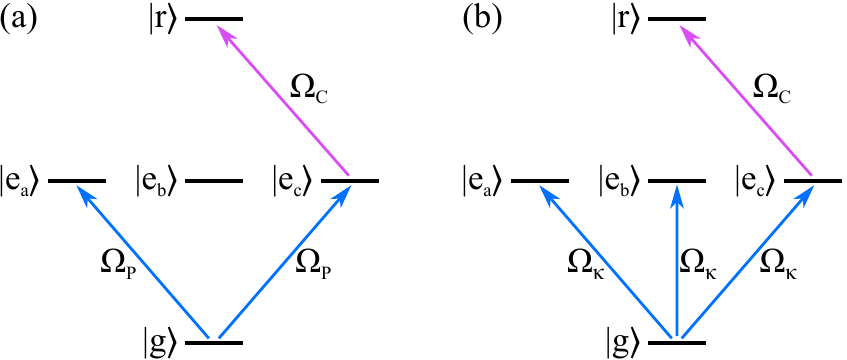}
\caption{(Color online) Energy level diagrams for: (a) laser excitation of Rydbergs, which form the narrow feature, (b) excitation of Rydbergs with rescattered field, which form the pedestal feature. \label{fig:figA}}
\end{figure}

We use the time-dependent optical Bloch equation (OBE) to model the non-interacting excitation dynamics of the atomic system. We consider a 5-level system with a ground state $|g\rangle$, three intermediate states $|e_{\mathrm{a}}\rangle$, $|e_{\mathrm{b}}\rangle$ and $|e_{\mathrm{c}}\rangle$, and one Rydberg excited state $|r\rangle$ (see Fig.~\ref{fig:figA}). 
The optical Bloch equation takes the form:
\begin{equation}
\dot{\rho} = \frac{i}{\hbar} \left[\rho,H\right] + L\left(\rho\right) + L_{\mathrm{d}}\left(\rho\right),
\end{equation}
where $\rho$ is the density matrix, $H$ is the Hamiltonian, and $L\left(\rho\right)$ and $L_{\mathrm{d}}\left(\rho\right)$ are the Lindblad operators used to include atomic state decay and additional decoherence due to finite laser linewidths respectively. An explanation of the implementation of this method for a three-level system can be found in \cite{PritchardThesis}. We have extended here this method to include the multiple intermediate states.

Due to the polarisation of the coupling beam and the angular momentum selection rules, only the intermediate state $|e_{\mathrm{c}}\rangle$ can be excited to $|r\rangle$. However, decay of the Rydberg state can occur to all three intermediate states with equal branching ratios and a lifetime $1/\Gamma_\mathrm{C}$. All three intermediate states can decay to the ground state with lifetime $1/\Gamma_\mathrm{P}$. These decay routes are included in the elements of the Lindblad operator $L\left(\rho\right)$. The diagonal terms of $L\left(\rho\right)$ take the form $\Gamma_\mathrm{P}\left(\rho_{e_a e_a}+\rho_{e_b e_b}+\rho_{e_c e_c}\right)$, $-\Gamma_\mathrm{P} \rho_{e_i e_i} + \left(\Gamma_\mathrm{C}/3\right) \rho_{rr}$ and $-\Gamma_\mathrm{C}\rho_{r r}$. The off-diagonal terms between $|r\rangle$ and $|e\rangle$ take the form $-\left(\Gamma_\mathrm{P} + \Gamma_\mathrm{C}/3\right)\rho_{r e_i}/2$, between $|r\rangle$ and $|g\rangle$, $-\Gamma_\mathrm{C} \rho_{r g}/2$, and between $|e\rangle$ and $|g\rangle$, $-\Gamma_\mathrm{P} \rho_{g e_i}/2$. Finally, the $|e_i\rangle$ to $|e_j\rangle$ off-diagonal terms take the form $-\Gamma_\mathrm{P} \rho_{e_i e_j}$.
The lifetime of the intermediate state is $1/\Gamma_{\mathrm{P}}$\ =\ 5\ ns \cite{Sansonetti2010}. The very small branching ratio to the $\mathrm{5}s\mathrm{4}d\,^{\mathrm{1}}D_{\mathrm{2}}\ $ state is negligible for the considered 5\ $\mu$s excitation time. We experimentally determine the lifetime of the Rydberg state to be $1/\Gamma_{\mathrm{C}}$\ =\ 69\ $\mu$s and in this model we assume the decay is directly to the intermediate states.

To model the narrow feature we used the excitation Rabi frequencies stated in the paper and treated the probe and coupling laser linewidths as fit parameters, which we varied to obtain the best fit to the low density spectrum in Fig.~2(a). The best fit was obtained with probe and coupling laser linewidths of  0\,MHz and $2\pi \times 1.6$\,MHz respectively. The same laser linewidths were used to model the time evolution of the laser-excited (narrow) component in Fig. \ref{fig:fig4} and in the propagation model (see below).

To model the pedestal feature we use the best fit laser linewidth for the coupling beam ($2\pi \times 1.6$\,MHz) and a broadband probe excitation field with linewidth equal to that of the intermediate state. The probe light is assumed to couple equally to all the sublevels of the intermediate state (see Fig.~\ref{fig:figA}), with the  Rabi frequency  $\Omega_{\kappa}$  treated as a fit parameter.


\begin{thebibliography}{99}
\bibitem{Walker1990}T.~Walker, D.~Sesko and C.~Wieman, Phys. Rev. Lett. {\bf 64}, 408 (1990); D.~Sesko, T.~Walker and C.~Wieman, J. Opt. Soc. Am. {\bf 8}, 946 (1991).
\bibitem{Labeyrie1999}G.~Labeyrie, F.~{de Tomasi}, J.~C.~Bernard, C.~A.~M\"uller, C.~Miniatura and R.~Kaiser, Phys. Rev. Lett. {\bf 83}, 5266 (1999).
\bibitem{Jonckheere2000}T. Jonckheere, C.~A~M\"uller, R.~Kaiser, C.~Miniatura, D.~Delande, Phys. Rev. Lett.  {\bf 85}, 4269 (2000).
\bibitem{Kupriyanov2005}D.~V,~Kupriyanov, Laser Phys. Lett. {\bf 3}, 223 (2006).
\bibitem{Baudouin2013}Q.~Baudouin, N.~Mercadier, V.~Guarrera, W.~Guerin and R.~Kaiser, Nature Physics {\bf 9}, 357 (2013).
\bibitem{Kong2014} C.~C.~Kwong, T.~Yang, M.~S.~Pramod, K.~Pandey, D.~Delande, R.~Pierrat and D.~Wilkowski, Phys. Rev. Lett. {\bf 113}, 223601 (2014).
\bibitem{Bidel2002}Y.~Bidel, B. Klappauf, J.~C.~Bernard, D.~Delande, G.~Labeyrie, C.~Miniatura, D.~Wilkowski and R.~Kaiser, Phys. Rev. Lett. {\bf 88}, 203902 (2002).
\bibitem{Dicke1954}R.~Dicke Phys.~Rev.~{\bf 93}, 99 (1954).
\bibitem{Rohlsberger2010}R.~R\"ohlsberger, K.~Schlage, B. Sahoo S.~Couet and R.~R\"uffer, Science {\bf 328}, 1248 (2010).
\bibitem{Keaveney2012}J.~Keaveney, A.~Sargsyan, U.~Krohn, I.~G.~Hughes, D.~Sarkisyan and C.~S.~Adams, Phys. Rev. Lett. {\bf 108}, 173601 (2012).
\bibitem{Pellegrino2014}J.~Pellegrino, R.~Bourgain, S.~Jennewein, Y.~R.~P.~Sortais, A.~Browaeys, S.~D.~Jenkins and J.~Ruostekoski, Phys. Rev. Lett. {\bf 113}, 133602 (2014).
\bibitem{Bromley2016}S.~L.~Bromley, B.~Zhu, M.~Bishof, X.~Zhang, T.~Bothwell, J.~Schachenmayer, T.~L.~Nicholson, R.~Kaiser, S.~F.~Yelin, M.~D.~Lukin, A.~M.~Rey and J.~Ye, Nat. Commun. {\bf 7}, 11039 (2016).
\bibitem{Olmos2013}B.~Olmos, D.~Yu, Y.~Singh, F.~Schreck, K.~Bongs and I.~Lesanovsky, Phys. Rev. Lett. {\bf 110}, 143602 (2013).
\bibitem{Zhu2015}B.~Zhu, J. Schachenmayer, M.~Xu, F.~Herrera, J.~G.~Restropo, M.~J.~Holland and A.~M.~Rey, New J. Phys. {\bf 17}, 083063 (2015).
\bibitem{Bettles2015}R.~J.~Bettles, S.~A.~Gardiner and C.~S.~Adams Phys. Rev. A {\bf 92}, 063822 (2015).
\bibitem{Pritchard2010}J.~D.~Pritchard, D.~Maxwell, A.~Gauguet, K.~J.~Weatherill, M.~P.~A.~Jones and C.~S.~Adams, Phys. Rev. Lett. {\bf 105}, 193603 (2010).
\bibitem{Lukin2001}M.~D.~Lukin, M.~Fleischhauer, R.~Cote, L.~M.~Duan, D.~Jaksch, J.~I.~Cirac and P.~Zoller, Phys. Rev. Lett. {\bf 87}, 037901 (2001).
\bibitem{Mohapatra2007}A.~K. Mohapatra, T.~R.~Jackson, and C. S. Adams, Phys. Rev. Lett. {\bf 98}, 113003 (2007).
\bibitem{Mauger2007}S.~Mauger, J.~Millen and M.~P.~A.~Jones, J.~Phys.~B {\bf 40}, F319 (2007).
\bibitem{Dudin2012} Y. O. Dudin and A. Kuzmich, Science {\bf 336}, 887 (2012).
\bibitem{Peyronel2012}T.~Peyronel, O.~Firstenberg, {Q.-Y.}~Liang, S.~Hofferberth, A.~V.~Gorshkov, T.~Pohl, M.~D.~Lukin and V.~Vuleti\'c, Nature {\bf 488}, 57 (2012)
\bibitem{Maxwell2013} D.~Maxwell, D.~J.~Szwer, D.~{Paredes-Barato}, H.~Busche, J.~D.~Pritchard, A.~Gauguet, K.~J.~Weatherill, M.~P.~A.~Jones and C.~S.~Adams, Phys. Rev. Lett. {\bf 110}, 103001 (2013).
\bibitem{Gorniaczyk2014} H. Gorniaczyk, C.~Tresp, J.~Schmidt, H.~Fedder and S.~Hofferberth, Phys. Rev. Lett. {\bf 113}, 053601 (2014).
\bibitem{Tiarks2014} D. Tiarks, S.~Baur, K.~Schneider, S.~D\"urr and G.~Rempe, Phys. Rev. Lett. {\bf 113}, 053602 (2014).
\bibitem{Gunter2012}G. G\"unter, M.~{Robert-de-Saint-Vincent}, H.~Schempp, C.~S.~Hoffman, S.~Whitlock and M.~Weidem\"uller, Phys. Rev. Lett. {\bf 108}, 013002 (2012).
\bibitem{Olmos2011}B. Olmos, W.~Li, S.~Hofferberth and I.~Lesanovsky, Phys. Rev. A {\bf 84}, 041607(R) (2011).
\bibitem{Friedler2005} I.~Friedler, D.~Petrosyan, M.~Fleischhauer and G.~Kurizki, Phys.~Rev.~A {\bf 72}, 043803 (2005)
\bibitem{Gorshkov2011}A.~V.~Gorshkov, J.~Otterbach, M.~Fleischhauer, T.~Pohl and M.~D.~Lukin, Phys. Rev. Lett. {\bf 107}, 133602 (2011).
\bibitem{Paredes2014} D.~{Paredes-Barato} and C.~S.~Adams, Phys. Rev. Lett. {\bf 112} 040501 (2014)
\bibitem{Khazali2015}M.~Khazali, K.~Heshami and C.~Simon, Phys. Rev. A {\bf 91} 030301(R) (2015)
\bibitem{Hofmann2013}C.~S.~Hofmann, G.~G\"unter, H.~Schempp, M. {Robert-de-Saint-Vincent}, M.~G\"arttner, J.~Evers, S.~Whitlock and M.~Weidem\"uller, Phys. Rev. Lett. {\bf 110}, 203601 (2013).
\bibitem{Garttner2013}M.~G\"arttner and J.~Evers, Phys. Rev. A {\bf 88}, 033417 (2013).
\bibitem{Boddy2014}D. Boddy, Doctoral Thesis, Durham University (2014), available at http://etheses.dur.ac.uk/10740/.
\bibitem{Wielandy1998}S.~Wielandy and A.~L.~Gaeta, Phys.~Rev.~A {\bf 58}, 2500 (1998).
\bibitem{Ates2006}C.~Ates, T.~Pohl, T.~Pattard and J.~M.~Rost, J.~Phys.~B {\bf 39}, L233 (2006). \bibitem{Ates2007}C.~Ates, T.~Pohl, T.~Pattard and J.~M.~Rost, Phys. Rev. A {\bf 76}, 013413 (2007).
\bibitem{Lesanovsky2013}I.~Lesanovsky and J.~P.~Garrahan, Phys.~Rev.~Lett. {\bf 111}, 215305 (2013).
\bibitem{Schempp2014}H. Schempp, G. G\"unter, M.~{Robert-de-Saint-Vincent},  C.~S.~Hoffman, D.~Breyel, A.~Komnik, D.~W.~Sch\"onleber, M.~G\''arttner, J.~Evers, S.~Whitlock and M.~Weidem\"uller, Phys. Rev. Lett. {\bf 112}, 013002 (2014).
\bibitem{Malossi2014}N.~Malossi, M.~M.~Valado, S.~Scotto, P.~Huillery, P.~Pillet, D.~Ciampini, E.~Arimondo and O.~Morsch, Phys.~Rev.~Lett. {\bf 113}, 023006 (2014).

\bibitem{Urvoy2015}A.~Urvoy, F.~Ripka, I.~Lesanovsky, D.~Booth, J.~P.~Schaffer, T.~Pfau and R.~L\"ow Phys.~Rev.~Lett. {\bf 114}, 203002 (2015).
\bibitem{Datsyuk2006}V.~M.~Datsyuk, I.~M.~Sokolov, D.~V.~Kupriyanov and M.~D.~Havey, Phys. Rev. A {\bf 74}, 043812 (2006).
\bibitem{Lochead2013}G.~Lochead, D.~Boddy, D.~P.~Sadler, C.~S.~Adams and M.~P.~A.~Jones, Phys. Rev. A {\bf 87}, 053409 (2013).
\bibitem{Millen2010}J.~Millen, G.~Lochead and M.~P.~A.~Jones, Phys. Rev. Lett. {\bf 105}, 213004 (2010).
\bibitem{Lochead2012}G.~Lochead, Doctoral Thesis, Durham University (2014), available at http://etheses.dur.ac.uk/6329/.
\bibitem{propnote}We have checked using our propagation model that propagation effects do not modify the shape over this range of $b$.
\bibitem{Labeyrie2003}G.~Labeyrie, E.~Vaujour, C.~A.~M\"uller, D.~Delande, C.~Miniatura, D.~Wilkowski and R. Kaiser, Phys. Rev. Lett. {\bf 91}, 223904 (2003).
\bibitem{Dalton1992}B.~J,~Dalton and P.~L.~Knight, Opt. Comm. {\bf 42}, 411 (1982).
\bibitem{Labeyrie2005}G.~Labeyrie, R.~Kaiser, D.~Delande, Appl. Phys. B {\bf 81}, 1001 (2005).
\bibitem{Pierrat2009}R.~Pierrat, B.~Gr{\'{e}}maud and D.~Delande, Phys. Rev. A {\bf 80}, 013831 (2009).
\bibitem{intennote}We also include a measured variation of the probe beam intensity with detuning  caused by the frequency response of the acousto-optic modulator used to swith the probe beam.
\bibitem{pedref}A Gaussian lineshape for the pedestal was used to increase the quality of the fits at low optical depth.
\bibitem{Tong2004}D.~Tong, S.~M.~Farooqi, J.~Stanojevic, S.~Krishnan, Y.~P.~Zhang, R.~Cot\'e, E.~E.~Eyler and P.~L.~Gould,  Phys.~Rev.~Lett. {\bf 93}, 063001 (2004).
\bibitem{Singer2004}K.~Singer, M.~{Reetz-Lamour}, T.~Amthor, L.~G.~Marcassa and M.~Weidem\"uller, Phys.~Rev.~Lett. {\bf 93}, 163001 (2004).

\bibitem{Olmos2014}B.~Olmos, I.~Lesanovsky and J.~P.~Garrahan, Phys. Rev. E {\bf 90}, 042147 (2014).
\bibitem{Marcuzzi2014} M. Marcuzzi, J. Schick, B. Olmos and I. Lesanovsky, J. Phys. A: Math. Theor. {\bf 47}, 482001 (2014).
\bibitem{Vaillant2012}C.~L.~Vaillant, M.~P.~A.~Jones and R.~M.~Potvliege, J. Phys. B {\bf 45}, 135004 (2012).
\bibitem{Killian1999}T.~C.~Killian, S.~Kulin, S.~D.~Bergeson, L.~A.~Orozco, C.~Orzel and S.~L.~Rolston,  Phys.~Rev.~Lett. {\bf 83}, 4776 (1999).

\bibitem{Avery1967}L.~W.~Avery and L.~L.~House, Astrophys. J. {\bf 152}, 493 (1968).
\bibitem{Ellinger1994}K.~Ellinger, J.~Cooper and P.~Zoller, Phys. Rev. A {\bf 49}, 3909 (1994).
\bibitem{Gremaud2006}B.~Gremaud, T.~Wellens, D.~Delande and C.~Miniatura, Phys. Rev. A {\bf 74}, 033808 (2006).
\bibitem{Pohl2006}T.~Pohl, G.~Labeyrie and R.~Kaiser, Phys. Rev. A {\bf 74}, 023409 (2006).
\bibitem{Wellens2010}T.~Wellens, T. Geiger, V.~Shatokin and A.~Buchleitner, Phys. Rev. A {\bf 82}, 013832 (2010).

\bibitem{Mendon2012}J.~T.~Mendon{\c c}a and R.~Kaiser, Phys. Rev. Lett. {\bf 108}, 033001 (2012).
\bibitem{Rodrigues2016}J.~D.~Rodrigues, J. A. Rodrigues, A.~V.~Ferreira, H.~Ter\c{c}as, R.~Kaiser and J.~T.~Mendon\c{c}a, arXiv:1604.08114 (2016).
\bibitem{PritchardThesis}J.~Pritchard, Doctoral Thesis, Durham University (2011).
\bibitem{Sansonetti2010}J.~E.~Sansonetti and G.~Nave, J. Phys. Chem. Ref. Data {\bf 39}, 033103 (2010). 


\end{thebibliography}
\end{document}